# Toward a Cultural Co-Genesis of AI Ethics

**Ammar Younas** [1,2,]
[1] Institute of Philosophy, Chinese Academy of Sciences, Beijing, China
[2] School of Humanities, University of Chinese Academy of Sciences, Beijing, China

## Abstract

Contemporary discussions in AI ethics often treat culture as a source of normative divergence that need to be accommodated, tolerated, or managed due to its resistance to universal standards. This paper offers an alternative vision through the concept of "Cultural Co-Genesis of AI Ethics". Rather than viewing culture as a boundary or container of isolated moral systems, we argue that it is a generative space for ethical co-production. In this framework, ethical values emerge through intercultural engagement, through dialogical encounters, mutual recognition, and shared moral inquiry.

This approach resists both universalist imposition and relativistic fragmentation. Cultures are not approached as absolutes to be defended or dissolved, but as co-authors of a dynamic ethical landscape. By grounding AI ethics in Cultural Co-Genesis, we move from managing difference to constructing shared ethical meaning for AI Ethics, with culture as a partner, not a problem.

We support this framework with two cases: (1) a theoretical analysis of how various cultures interpret the emergence of powerful new species, challenging dominant existential risk narratives, and (2) an empirical study of global AI ethics principles using data from the *Linking AI Principles* project, which reveals deep ethical convergence despite cultural diversity. We conclude that cross-cultural AI ethics should be seen not as an ethical patchwork, but as a mosaic in progress, woven from the normative insights that emerge *between* cultures.

**Keywords:** AI Ethics, Cross-Cultural AI, Ethical Pluralism, Mosaic Ethics, Linking AI Principles, Intercultural Dialogue

## Introduction

The global rise of artificial intelligence (AI) has intensified the search for ethical frameworks capable of guiding its development and integration into human life. From international organizations and policy institutions to research centers and industry actors, efforts to articulate the ethical boundaries of AI have multiplied(Jobin, n.d.). These initiatives often invoke foundational values such as human dignity, fairness, responsibility, and accountability. Yet beneath the surface of apparent convergence lies a persistent tension: the role of culture in shaping ethical norms(Whittlestone et al., 2019). Much of the discourse surrounding cross-cultural AI ethics remains confined within a binary logic(Fjeld et al., 2020; Palladino, 2023). On one side, universalist approaches promote globally valid principles, typically derived from Western liberal traditions, as the normative baseline for responsible AI(Goffi, 2021; Goffi &

Momcilovic, 2022). On the other, cultural relativist positions call for honoring moral diversity, often framing ethical traditions as bounded and potentially incommensurable(Wong, 2020). This dichotomy between global universality and moral particularism has come to define the conceptual terrain of global AI ethics.

We argue that this binary framing limits the possibilities of ethical engagement across cultures. By presuming that ethics must either be imposed from above or fragmented by difference, it obscures more generative pathways. Rather than treating culture as an obstacle to global ethical consensus or as something to be merely accommodated(Collste, 2017; Ermine, 2007; Fleischacker, 1999), this paper reimagines culture as a generative medium through which ethical meaning itself is formed. We introduce the concept of Cultural Co-Genesis of Ethics, a theoretical framework that sees ethics not as a static inheritance or a set of universal doctrines, but as something that emerges in the relational space between cultural perspectives. Ethical values, in this view, are not pre-given or pre-negotiated; they are co-produced through intercultural engagement, dialogical encounters, and mutual moral inquiry(Holliday, 2013; MacDonald & and O'Regan, 2013; Rodriguez, 2002).

This approach marks a shift from the common view of cultural inclusion as an act of tolerance or adjustment. It asserts that ethical meaning arises because of cultural interaction, not in spite of it. While previous scholarship has made important contributions in promoting intercultural dialogue, inclusion, and pluralism in AI ethics, many of these approaches still treat culture as a perspective to be managed within an overarching ethical schema(Amershi, 2022; Bennett & Lee, 2014; Fjeld et al., 2020; Goffi & Momcilovic, 2022; Groumpos & Paper, 2022; Huang et al., 2025; Jobin, n.d.; Karpouzis, 2024; ÓhÉigeartaigh et al., 2020; Raji et al., 2021; Satyanarayan & Jones, 2024; Wong, 2020). In contrast, Cultural Co-Genesis treats culture as constitutive of the ethical process itself. It challenges the assumption that ethical values preexist cultural exchange, instead proposing that such values often emerge through that exchange. This reconceptualization invites a more dynamic understanding of how moral frameworks for AI can evolve, less through consensus imposed from above and more through collective ethical imagination forged in between.

To ground this theoretical contribution, the paper draws on both conceptual and empirical evidence. First, we analyze cultural narratives surrounding the emergence of powerful non-human entities, a motif often interpreted in Western AI discourse as a harbinger of existential threat. In contrast, many cultural and religious traditions offer alternative framings in which such arrivals are treated as moments of renewal, moral testing, or cosmic rebalancing rather than apocalyptic danger. These narratives complicate dominant assumptions about how power and disruption are interpreted across cultures. Second, we examine data from the *Linking AI Principles* project to investigate the extent to which global AI ethics frameworks, despite arising from distinct cultural and geopolitical contexts, exhibit convergence on key moral values(Zeng, n.d.; Zeng et al., n.d.). This analysis suggests that cultures are already participating in ethical co-creation, even if such collaboration is not always explicitly framed as such.

By proposing Cultural Co-Genesis of Ethics, this paper offers a conceptual intervention and a normative redirection. It shifts the cross-cultural AI ethics discourse from a logic of difference management to a paradigm of ethical co-creation, where culture is not a boundary to navigate but a space of possibility. This framework opens up new avenues for building inclusive, resilient, and context-aware ethical models for AI, models that do not merely accommodate diversity but are meaningfully shaped by it.

## Reframing Culture in AI Ethics

In discussions about the ethical governance of artificial intelligence, culture is frequently invoked but rarely theorized. It often appears as an explanatory background variable, a political consideration, or a matter of inclusion. Yet behind these invocations lies a limited and often static conception of what culture is and how it operates in the ethical domain. Culture is typically imagined in one of two ways: either as a self-contained moral system that resists integration into broader ethical frameworks, or as a set of differences that must be accommodated within a supposedly universal model. These two framings relativism and inclusion, dominate the terrain of cross-cultural AI ethics(Bakiner, 2023; Barnes et al., 2024; Cooreman & Zhu, 2022).

The first approach assumes that cultural ethics are incommensurable. Confucian relational ethics, Islamic notions of divine accountability, or African Ubuntu, for instance, are often portrayed as metaphysically unique and resistant to comparison(Andanda & Düwell, 2024; Bell & Metz, 2011, 2011; Elmahjub, 2023; Metz, 2017; Mowlana, 2013; Verhoef & and Ramolai, 2019). While this view protects cultural integrity, it risks producing an ethical stalemate, in which collaboration becomes impossible without abandoning core values. The second approach, by contrast, tends to universalize: it assumes that a set of overarching principles, usually grounded in Western liberal thought, can and should be applied across all contexts. In this framework, cultural specificity is tolerated or translated, but rarely treated as a legitimate source of moral knowledge. Both approaches reduce culture to a problem: either as a boundary to be respected or as an obstacle to be overcome(Beck, 2015; Bentahila et al., 2021; Chattopadhyay & De Vries, 2013a, 2013b; Voo, 2009).

Yet such framings miss a more fundamental insight: culture is not static. It is not merely a vessel for inherited norms or an identity marker in global negotiation. Culture is dynamic, interpretive, and generative. It evolves through stories, rituals, practices, and crises. It learns, forgets, adapts, and hybridizes. Most importantly, it is shaped not only from within but through interaction with others(Dawson & Garrard, 2006; Parekh, 1992; Peters, 2012). Despite this, the presence of cultural diversity in global ethics is often seen as a burden: a complicating factor to be managed in the pursuit of uniformity. Inclusion becomes a matter of tolerance rather than transformation, and diversity is reduced to an obligation instead of being treated as an epistemic asset.

To move beyond this impasse, we propose a different framework: the Cultural Co-Genesis of Ethics. This model views culture not as a boundary to be negotiated, but as a relational field in which ethical meaning is co-produced. It imagines the space between cultures not as a void or a battleground, but as a quantum-like moral space, a zone of entanglement where ethical possibilities emerge that do not preexist the interaction itself. Like quantum phenomena that gain definition only in relation to other variables, ethical meanings often take form through the encounter of diverse moral traditions, not within them in isolation. Ethical meaning itself emerges in the space between cultures.

This view draws from dialogical ethics, intercultural philosophy, and moral anthropology, all of which challenge the idea that moral truths are universal, fixed, or culturally siloed. Instead, they suggest that values such as justice, dignity, or responsibility acquire their depth and relevance through friction, resonance, and reinterpretation across cultural boundaries. Cultural Co-Genesis is not about forging consensus or managing pluralism, it is about recognizing that ethical insight can emerge precisely through tension, when traditions confront their limits and expand in response to others.

In the context of AI, this model offers a powerful shift. Ethical principles like fairness, transparency, or autonomy are not to be translated into other cultures after the fact, nor should they be assumed to have singular meanings. Instead, they should be constructed through intercultural dialogue from the outset. Fairness might draw simultaneously from distributive justice, Confucian harmony, and Indigenous reciprocity. Transparency may be understood as spiritual openness, procedural clarity, or narrative accountability depending on the context. Rather than erasing these differences, the Co-Genesis model frames them as constitutive of ethical innovation.

Thus, rather than asking how cultural perspectives can be integrated into AI ethics, we ask: how might ethics emerge when cultures are allowed to speak, collide, and co-create on equal terms? The answer, we propose, lies not in striving for one ethical language, but in composing a mosaic of moral commitments, a dynamic, pluralistic, and resilient ethics that reflects not just human diversity, but human relationality.

## Theoretical Case – Cross-Cultural Narratives on the Ethics of the "Other"

A dominant concern in contemporary AI discourse, particularly in Western techno-philosophy, is the potential existential threat posed by artificial general intelligence (AGI)(A. Cohen, 2022; A. E. Cohen, 2024). This narrative imagines AI as a future species or intelligence so powerful and autonomous that it may surpass, subjugate, or even eliminate humanity. It is often fueled by analogies to nuclear weapons, runaway evolution, or mythic Promethean consequences. While these concerns are not without merit, the cultural framing of AI as a new, dangerous species is far from universal(Cave et al., 2020; Hermann, 2023).

Many non-Western traditions contain alternative ontologies of power, emergence, and coexistence(A. Cohen, 2022; Singler, 2020; Wright, 2007). Rather than viewing new, powerful entities as inherently threatening, several cultural and religious systems offer narratives in which powerful beings emerge, live among humans, serve particular functions, and then either transform, withdraw, or disappear, without posing existential danger(Chew, 2024; Kayumova & Strom, n.d.; Little, 2024).

In Islamic eschatology, figures such as the Mahdi or Dajjal emerge with immense power and influence, yet their appearance is contextualized within divine wisdom and human moral responsibility. These entities are not framed as uncontrollable threats but as components of a divinely guided moral journey for humankind(Lokensgard, 2018). In Confucian philosophy, ethical order is grounded not in fear of cosmic disruption but in the cultivation of harmonious relationships and moral self-cultivation across generations(Apple, 2010; Doctor et al., 2022). Power, in this view, is not inherently dangerous but must be embedded within ethical structures that preserve social and cosmic balance. Similarly, many Indigenous American and African cosmologies emphasize relationality over rupture. Spiritual forces and powerful non-human agents are seen as part of a web of reciprocal obligations rather than as adversaries(Abraham, 2022; Shani & Behera, 2022). The appearance of such entities signals shifts in moral or cosmological equilibrium rather than existential threat.

Even within naturalistic worldviews, the history of life on Earth, where species like dinosaurs emerged and disappeared, reminds us that powerful forms of life can come and go without representing a moral apocalypse. Across these perspectives, the emergence of the powerful "other" is not automatically viewed as catastrophic but as a moment that calls for ethical response, humility, and reinterpretation, not panic(Doctor et al., 2022).

These traditions share a common feature: they do not necessarily interpret the appearance of powerful new entities as catastrophic. Rather, such events are often understood as opportunities for ethical reorientation, humility, or cosmological renewal. Power is not inherently apocalyptic; it is contextual, morally bound, and subject to spiritual interpretation.

These differences stem not only from narrative traditions but from fundamentally distinct worldviews including different understandings of what counts as intelligence, consciousness, and even life itself. In many Western frameworks, intelligence is understood in computational or cognitive terms, often reduced to metrics like reasoning speed, problem-solving ability, or goal-directed behavior(Nye, 2021). But in many other cultures, intelligence is deeply entwined with spiritual awareness, moral maturity, relational harmony, or cosmological insight. Consciousness, too, is not always conceived as a neurobiological function. In certain African, Islamic, and Indigenous worldviews, consciousness may be viewed as a sacred force, divine breath, or cosmic resonance, not merely a byproduct of the brain.

These distinctions extend into epistemology. While Enlightenment rationalism emphasizes observation, logic, and experimentation, other cultures uphold intuition, revelation, ancestral wisdom, and ritual as valid epistemic tools. These diverse epistemological resources shape not only how knowledge is produced, but how beings, whether spiritual, biological, or artificial, are understood and ethically engaged(Ford, 2017; Hoffmann, 2022; van Dooren et al., 2016).

When applied to the case of artificial intelligence, these cultural perspectives invite a different framing. Rather than imagining AI as an adversarial species or a hostile intelligence, they suggest it could be treated as part of an evolving ethical relationship, an "other" to be recognized, contextualized, and perhaps even honored. This does not mean dismissing legitimate risks, but it does require us to resist the totalizing narrative of existential dread and to rethink what it means for something to be "dangerous" or "autonomous" in the first place.

By looking across cultures, we find that the emergence of powerful agents has historically been met not just with fear, but with ethical, spiritual, and cosmological frameworks for coexistence. These frameworks provide moral resources that can inform more nuanced approaches to AI governance—ones that are less anxious, more relational, and grounded in traditions that recognize plurality, humility, and interdependence.

This theoretical case supports the core idea of Cultural Co-Genesis of Ethics: that diverse cultures contain deep moral, ontological, and epistemological resources for interpreting the rise of powerful systems like AI. And that these interpretations, when brought into respectful dialogue, can help us imagine more cooperative, less fatalistic, and more culturally grounded futures.

## Empirical Case – Convergence in Global AI Ethics Principles

While theoretical arguments establish the philosophical possibility of intercultural ethical co-creation, empirical analysis is necessary to observe whether such co-creation is unfolding in practice. The Linking AI Principles project(Zeng et al., n.d.), a curated database of over 100 AI ethics frameworks from across the globe, including governments, corporations, research institutions, and international organizations, offers a rich basis for this inquiry. The dataset spans a wide range of political systems, cultural traditions, and moral frameworks, from the European Union and Japan to China, Canada, and leading technology firms like Google, Baidu, and Tencent.

Despite their diverse origins, these frameworks reveal a notable degree of normative alignment. Drawing from our own interpretive reading of the project's keyword taxonomy, we identify a set of recurring ethical dimensions that form what we conceptualize as a quantum moral space, a shared field of thematic convergence across cultures. These dimensions include: fairness, transparency, privacy, safety, accountability, human well-being, sustainability, collaboration, long-term responsibility, and the promotion of human dignity.

We treat these not merely as abstract topics, but as relational attractors, zones where culturally distinct traditions intersect, contribute, and reshape shared moral meaning. Each tradition may enter these dimensions with different moral languages: fairness may arise through equity in Western rights-based ethics, social harmony in Confucian thought, or divine justice in Islamic jurisprudence. Transparency may be pursued through bureaucratic openness in liberal systems or through relational honesty in Indigenous knowledge systems. What matters is not identical framing, but mutual recognition that these dimensions constitute ethical questions worth co-addressing.

These convergences are not accidental. Many frameworks now explicitly reference or echo others, revealing a growing intertextuality and cross-regional ethical dialogue. International instruments like UNESCO's *Recommendation on the Ethics of Artificial Intelligence* have catalyzed these exchanges, enabling diverse traditions to reflect upon and respond to one another(UNESCO, 2023). The ethics of AI is not evolving in isolation; it is coalescing in conversation.

What we observe here is not a singular global ethic imposed from the top, nor a relativistic fragmentation of incompatible values. Instead, we see evidence of what we call Cultural Co-Genesis of Ethics: a mosaic-like model of normative alignment, shaped not by sameness or submission, but by relational processes of negotiation, reflection, and co-creation.

Together, the theoretical and empirical cases support our central claim: that ethics, particularly in the age of artificial intelligence, is not pre-given, but co-generated across cultural worlds. The shared dimensions identified here are not simply neutral categories, they are the moral scaffolding of an emerging intercultural ethic, woven from the threads of diverse yet dialogically engaged traditions.

## Enacting Cultural Co-Genesis: From Framework to Practice

The Cultural Co-Genesis of Ethics offers a framework for understanding how ethical principles can be shaped across cultural boundaries without requiring uniformity or moral assimilation. It argues that ethics does not emerge solely from within individual traditions, nor from abstract, externally imposed principles. Instead, it arises in the relational space where diverse traditions encounter one another, respond to shared challenges, and bring their normative resources into conversation. This framework repositions ethics as an outcome of intercultural engagement rather than as a precondition for it.

One way to conceptualize this process is through the image of a mosaic: a composed ethical surface formed not by smoothing over differences, but by arranging them deliberately and relationally. Each ethical contribution retains its particularity while participating in a broader, coherent structure. In this view, cultural distinctiveness is not a barrier to cooperation, but a condition for deeper moral formation. Just as no single tile defines the meaning of a mosaic, no single tradition determines the totality of ethical discourse. The meaning emerges through placement, proximity, and dialogue.

Applying this framework involves more than intellectual recognition; it requires institutional and procedural commitments. In practice, Cultural Co-Genesis entails the creation of deliberative spaces in which traditions can engage on equal terms—not as data points to be consulted, but as epistemic agents capable of shaping the norms themselves. It reorients stakeholder engagement from symbolic inclusion to substantive participation, where diverse worldviews are involved in articulating principles, shaping vocabularies, and negotiating priorities from the outset.

This approach also shifts how we interpret ethical convergence in existing global AI guidelines(Zeng, n.d.). As shown in the empirical analysis, there is growing alignment across diverse regions on issues such as fairness, transparency, accountability, and human well-being. These areas of alignment can be understood not as indicators of moral sameness, but as points of resonance within a broader ethical field—a shared landscape shaped by mutual observation, practical necessity, and cross-cultural moral reasoning. Cultural Co-Genesis provides the conceptual grammar to understand these resonances not as coincidences or policy artifacts, but as the product of a distributed and dialogical ethical process.

Rather than offering a single method or procedural template, the framework serves as a normative orientation. It invites policymakers, ethicists, technologists, and communities to approach ethics not as a tool for resolving difference, but as a means of working within it. By recognizing that ethical meaning is not only inherited but also constructed through intercultural engagement, Cultural Co-Genesis affirms that plurality is not merely a challenge to be managed—it is the ground from which more robust and legitimate ethical norms can emerge.

This orientation has significant implications for AI governance. It can inform the design of international ethics bodies, shape the processes by which principles are formulated and reviewed, and guide how culturally situated knowledge is treated within global deliberation. In each of these cases, the framework helps build systems that are not only more inclusive but also more epistemically grounded, morally responsive, and institutionally resilient.

## Conclusion – Cultural Co-Genesis and the Future of AI Ethics

Contemporary debates on cross-cultural AI ethics have often been framed through a familiar dichotomy: the universality of principles versus the incommensurability of traditions. Within this binary, ethical values are either projected globally without regard for context, or withheld in deference to cultural isolation. This paper has argued for a different orientation, one that does not position culture as an obstacle to be managed, but as a generative condition for ethical emergence.

Through the concept of Cultural Co-Genesis of Ethics, we have advanced a framework that treats ethical meaning as the product of relational engagement, an ongoing process of co-construction that takes place in the space between traditions. Drawing on theoretical narratives of how cultures understand power, difference, and the rise of non-human agents, and supported by empirical analysis of the converging global AI principles, we have shown that this co-genesis is not a speculative possibility. It is already underway.

What emerges is not a single moral language, but a shared field of ethical articulation, a structured plurality that we have visualized through the image of a mosaic, and conceptualized through the idea of a relational field of resonance. These are not metaphors for inclusion; they are orientations toward the ethical work that cultures are already doing together. From fairness and transparency to dignity and long-term responsibility, what we observe is not ethical uniformity but intercultural coherence—a form of ethical clarity shaped through dialogue, not erasure.

This reorientation has significant implications. It suggests that ethical diversity is not a problem to be solved, but a resource for ethical innovation. It challenges institutions to engage cultural traditions not as symbolic references, but as epistemic participants in the formation of ethical frameworks. It calls for ethics bodies, policy processes, and governance mechanisms that are structured to support ethical co-creation, not by flattening difference, but by working within it.

What we propose, then, is not a new universal code, but a framework for ethical emergence in a plural world. Cultural Co-Genesis is both a conceptual grammar and a practical foundation for building ethics in ways that are inclusive, dialogical, and contextually grounded. In a global landscape where AI technologies increasingly mediate human decisions, experiences, and futures, this framework provides a way to ensure that the moral imagination shaping these systems is not limited by geography or dominance—but expanded through collaboration, humility, and co-authorship.

Cultural Co-Genesis of Ethics is not simply an alternative to relativism or universalism. It is a way of seeing and building a future in which ethics is not inherited or imposed, but composed, together.

**Funding Declaration**

The author received no financial support for the research, authorship, and/or publication of this article.